\documentclass[twocolumn,showpacs,preprintnumbers,amsmath,amssymb]{revtex4}

\usepackage{graphicx}
\usepackage{dcolumn}
\usepackage{bm}
\begin{document}

\title{Continuous-time quantum walks on Erd\"{o}s-R\'{e}nyi networks}
\author{X.P. Xu$^{1,2}$}
\author{F. Liu$^{2}$}
\affiliation{%
$^1$Institute of High Energy Physics, Chinese Academy of Science,
Beijing 100049, China \\
$^2$Institute of Particle Physics, HuaZhong Normal University, Wuhan
430079, China
}%

\date{\today}

\begin{abstract}
We study the coherent exciton transport of continuous-time quantum
walks (CTQWs) on Erd\"{o}s-R\'{e}nyi networks. The
Erd\"{o}s-R\'{e}nyi network of $N$ nodes is constructed by
connecting every pair of nodes with probability $p$. We numerically
calculate the ensemble averaged transition probability of quantum
transport between two nodes of the networks. For finite networks, we
find that the limiting transition probability is reached very
quickly. For infinite networks whose spectral density follows the
semicircle law, the efficiencies of the classical and
quantum-mechanical transport are compared on networks of different
average degree $\bar{k}$. In the long time limiting, we consider the
distribution of the ensemble averaged transition probabilities, and
show that there is a high probability to find the exciton at the
initial node. Such high return probability almost do not alter in a
wide range of connection probability $p$ but increases rapidly when
the network approaches to be fully connected. For networks whose
topology is not extremely connected, the return probability is
inversely proportional to the network size $N$. Furthermore, the
transport dynamics are compared with that on a random graph model in
which the degree of each node equals to the average degree $\bar{k}$
of the Erd\"{o}s-R\'{e}nyi networks.
\end{abstract}
\pacs{05.60.Gg, 03.67.-a, 05.40.-a}
\maketitle
\section{Introduction}
During the last few years, the coherent exciton dynamics in quantum
system has been extensively studied by both experimental and
theoretical methods~\cite{rn1,rn2,rn3,rn4}. The dynamical behavior
of such process depends on the underlying structure of the system
under study. Most of previous studies on coherent exciton dynamics
are based on simple structures, for example, the
line~\cite{rn5,rn6}, cycle~\cite{rn7,rn8}, hypercube~\cite{rn9},
Cayley tree~\cite{rn10}, dendrimers~\cite{rn11},
polymers~\cite{rn12} and other regular networks with simple
topology. To the best of our knowledge, the dynamics of exciton on
random network have not received much attention~\cite{rn13}.

In this paper, we consider the coherent exciton transport on random
networks of Erd\"{o}s-R\'{e}nyi (ER). The coherent exciton dynamics
is modeled by continuous-time quantum walks (CTQWs), which is a
quantum version of the classical random walk and widely studied by
various researchers to describe the relaxation processes in complex
systems~\cite{rn14,rn15}. In the mathematical literature, the ER
random network is defined as follows~\cite{rn16,rn17,rn18}: Starting
with $N$ disconnected nodes, every pair of nodes is connected with
probability $p$ ($0<p<1$) and multiple connections are prohibited.
The ER random network is one of the oldest and best studied models
of networks, and possesses the considerable advantage of being
exactly solvable for many of its average properties in the limit of
large network size~\cite{rn19}. For instance, one interesting
feature, which was demonstrated in their original papers, is that
the model shows a phase transition with increasing $p$ at which a
giant component forms~\cite{rn19,rn20}. An alternative and
equivalent representation of the ER random graph is to express the
graph not in terms of $p$ but in terms of the average degree
$\bar{k}$ of the nodes, which is related to the connection
probability $p$ as: $\bar{k}=p(N-1)\approx pN$, where the last
approximate equality is hold for large $N$.

In the limit of large network size $N$, the degrees of ER random
network follow a Poisson distribution peaked at the average degree
$\bar{k}$. In order to contrast the resemblance and difference of
the transport dynamics on networks with the same average degree, we
consider the coherent exciton transport on a configuration model of
random networks in which the degree $k$ of each node equals to the
average degree $\bar{k}$ ($\bar{k} \in Integers$) of the
Erd\"{o}s-R\'{e}nyi networks. The method for generating the graph is
as follows~\cite{rn21}: one assigns each node $\bar{k}$ ($\bar{k}
\in Integers$) "stubs" -ends of edges emerging from the nodes, and
then one chooses pairs of these stubs uniformly at random and joins
them together to make complete edges. When all stubs have been used
up, the resulting graph is a random member of the ensemble of graphs
with the equal degree~\cite{rn21,rn19}. The configuration model of
random networks can also be implemented by using the
edge-interchanging algorithm, which randomly interchange two
existing edges while keep the degree sequence
unchanged~\cite{rn22,rn23}. The configuration model is one of the
most successful algorithms proposed for network formation, and has
been extensively used as a null model in contraposition to real
networks with the same degree distribution in biology, robustness,
epidemics spreading and other dynamical processes taking place on
complex networks~\cite{rn22,rn24,rn25}. Here, we adopt this idea to
compare the transport behavior on the two network models. As we will
show, although the ensembles of ER model and configuration model
have the same average number of connections, the transport dynamics
on the two models are different.

The paper is structured as follows: In the next section, we briefly
review the classical and quantum transport on networks presented in
Refs.~\cite{rn26,rn27}. In Sec.~\uppercase\expandafter{\romannumeral
3} we study the time evolution of the ensemble averaged return
probability on ER networks with different parameters.
Section~\uppercase\expandafter{\romannumeral 4} presents the
efficiencies of the classical and quantum mechanical transport, and
try to reveal how the model parameter affects the transport
efficiency. In Sec.~\uppercase\expandafter{\romannumeral 5}, we
consider the distribution of the long time averaged transition
probabilities, and explore how the average return probability is
related to network parameters. In
Sec.~\uppercase\expandafter{\romannumeral 6}, we consider the
transport dynamics on extremely connected networks. Conclusions and
discussions are given in the last part,
Sec.~\uppercase\expandafter{\romannumeral 7}.
\section{transport on networks}
The coherent exciton dynamics on a connected network is modeled by
the continuous-time quantum walks (CTQWs), which is obtained by
replacing the Hamiltonian of the system by the classical transfer
matrix, $H=-T$~\cite{rn28}. The transfer matrix $T$ relates to the
Laplace matrix by $T= -\gamma A$~\cite{rn10}. Here, for the sake of
simplicity, we assume the transmission rate $\gamma$ for all
connections equals to $1$. The Laplace matrix $A$ has nondiagonal
elements $A_{ij}$ equal to $-1$ if nodes $i$ and $j$ are connected
and $0$ otherwise. The diagonal elements $A_{ii}$ equal to the
number of total links connected to node $i$, i.e., $A_{ii}$ equals
to the degree of node $i$. The states $|j>$ endowed with the nodes
$j$ of the network form a complete, ortho-normalised basis set,
which span the whole accessible Hilbert space, i.e., $\sum_k
|k><k|=1$, $<k|j>=\delta_{kj}$. The transport processes are governed
by the master equation or Schr\"{o}dinger equation~\cite{rn10}. The
classical and quantum mechanical transition probabilities to go from
the state $|j>$ at time $0$ to the state $|k>$ at time $t$ are given
by $p_{k,j}(t)=<k|e^{-tA}|j>$ and $\pi_{k,j}(t)=|\alpha_{k,j}(t)|^2=
|<k|e^{-itH}|j>|^2$~\cite{rn10}, respectively. Generally speaking,
to calculate the transition probabilities, all the eigenvalues and
eigenvectors of the transfer operator and Hamiltonian are required.
We use $E_n$ to represent the $n$th eigenvalue of $H$ and denote the
orthonormalized eigenstate of Hamiltonian by $|q_n>$, such that
$\sum_n|q_n><q_n|=1$. The classical and quantum transition
probabilities between two nodes can be written as,
\begin{equation}\label{eq1}
p_{k,j}(t)=\sum_n e^{-tE_n}<k|q_n><q_n|j>,
\end{equation}
and
\begin{equation}\label{eq2}
\begin{array}{ll}
\pi_{k,j}(t)&=|\alpha_{k,j}(t)|^2 \\
&=\sum_{n,l} e^{-it(E_n-E_l)}\\
&~~\times <k|q_n><q_n|j><k|q_l><q_l|j>.
\end{array}
\end{equation}
The above equations give the general expressions of the classical
and quantum transition probabilities, which explicitly depends on
the eigenvalues and eigenvectors of the transfer matrix or
Hamiltonian. A particular feature related to the transport is the
return probability, which is the probability of finding the exciton
at the initial node. The transition probability depends on the
specific topology of the generated single network, therefore it is
appropriate to consider its ensemble averages.
\section{Averaged return probabilities}
The average of the classical and quantum return probabilities
$p_{j,j}(t)$ and $\pi_{j,j}(t)$ over all nodes of the network are
\begin{equation}\label{eq3}
\begin{array}{ll}
\bar{p}(t)&=\frac{1}{N}\sum_n e^{-tE_n}\sum_j<q_n|j><j|q_n> \\
&=\frac{1}{N}\sum_n e^{-tE_n},
\end{array}
\end{equation}
and
\begin{equation}\label{eq4}
\begin{array}{ll}
\bar{\pi}(t)&=\frac{1}{N}\sum_j \pi_{j,j}(t)=\frac{1}{N}\sum_j |\alpha_{j,j}(t)|^2 \\
&=\frac{1}{N}\sum_{n,l} e^{-it(E_n-E_l)}\\
&~~\times \sum_j <j|q_n><q_n|j><j|q_l><q_l|j>.
\end{array}
\end{equation}
The classical $\bar{p}(t)$ is only dependent on the eigenvalues and
decays monotonically from $\bar{p}(0)=1$ to the equipartition
$\lim_{t\rightarrow \infty}\bar{p}(t)=1/N$. The quantum
$\bar{\pi}(t)$ is dependent on the eigenvalues and eigenvectors,
which is cumbersome in the numerical calculations. The above
equations present the average of return probabilities over all nodes
on a specific single network. In order to reduce the statistical
fluctuation, we further average the return probabilities over
distinct single networks, i.e.,
\begin{equation}\label{eq5}
<\bar{p}(t)>=\frac{1}{R}\sum_{r=1}^R \bar{p}^r(t),
\end{equation}
and
\begin{equation}\label{eq6}
<\bar{\pi}(t)>=\frac{1}{R}\sum_{r=1}^R \bar{\pi}^r(t),
\end{equation}
where the index $r$ denotes the $r$th generated ER network.
Throughout this paper, we denote the average over network nodes by a
bar (e.g., $\bar{k}$, $\bar{p}(t)$, $\bar{\pi}(t)$, etc.), and the
average over different realizations by a bracket (e.g.,
$<\bar{p}(t)>$) while the actual values by undecorated characters.
\begin{figure}
\scalebox{1.2}[1.2]{\includegraphics{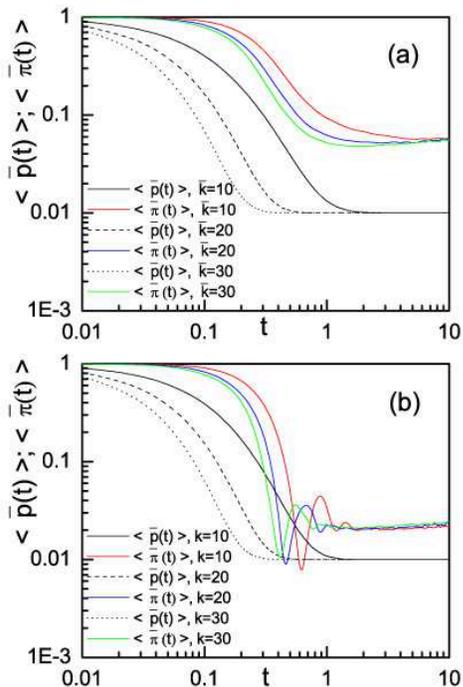}} \caption{(Color
online) Averaged return probabilities $<\bar{p}(t)>$ and
$<\bar{\pi}(t)>$ for random networks generated by the ER model (a)
and configuration model (b) with different values of degree. The
networks are of size $N=100$ and the (average) degree of the
networks are $\bar{k}=10$, $\bar{k}=20$ and $\bar{k}=30$ (see the
different line styles), respectively. All the curves are averaged
over $100$ independent realizations.
 \label{fg1}}
\end{figure}

Figure~\ref{fg1}(a) shows the ensemble averaged return probabilities
$<\bar{p}(t)>$ and $<\bar{\pi}(t)>$ on ER networks of size $N=100$
with average degree $\bar{k}=10$, 20 and 30. For classical transport
$<\bar{p}(t)>$ reaches the equipartition $\lim_{t\rightarrow
\infty}\bar{p}(t)=1/N$ very quickly. The curves at intermediate
times follow stretched exponential decay, which differs from power
law decay ($t^{-0.5}$) for the cycle graph~\cite{rn29}. The
exponential decay of $<\bar{p}(t)>$ indicates that a classical
excitation will quickly spread the whole network and occupy each
node with an uniform probability $1/N$ in a short time. It is
evident that the excitation reaches the equipartition $1/N$ more
quickly on networks with more connections (compare the curves in
Fig.~\ref{fg1}(a)). For quantum transport $<\bar{\pi}(t)>$ also
decays quickly in the intermediate times and then reach a final
plateau. This plateau is larger than the equip-partitioned
probability $1/N$. After a careful examination, we find such plateau
corresponds to a constant value $<\bar{\pi}(t)>\approx 0.065\pm
0.01$. Increasing the average degree $\bar{k}$ nearly does not
change this value (compare the curves in Fig.~\ref{fg1}(a)). We note
that here $<\bar{\pi}(t)>$ is smooth and does not display the
oscillatory behavior, in contrast to the case for the cycle graph in
which the return probability shows a quasi-periodic
pattern~\cite{rn29}. The non-periodic behavior of $<\bar{\pi}(t)>$
may be attributed to the large connectivity of the networks consider
here.

In Fig.~\ref{fg1}(b), we show the same plot of $<\bar{p}(t)>$ and
$<\bar{\pi}(t)>$ on a random graph model in which the degrees of
each node are exactly equal to $k=10$, 20 and 30. The behavior of
$<\bar{p}(t)>$ is almost the same as that on the ER networks.
However, $<\bar{\pi}(t)>$ is quite different. $<\bar{\pi}(t)>$
oscillates at intermediate times and also reaches a constant value.
Such constant value ($0.028\pm 0.003$) is lower than that of the ER
networks but larger than that of the cycle graph
($(2N-2)/N^2=0.0198$ for even-numbered networks and
$(2N-1)/N^2=0.0199$ for odd-numbered networks)~\cite{rn30}. The
random networks generated by the ER model and configuration model
have the same number of total connections (on average), but the
average return probabilities are quite different. Such difference
may be caused by the different ensembles generated by the network
models.
\begin{figure}
\scalebox{1.6}[1.6]{\includegraphics{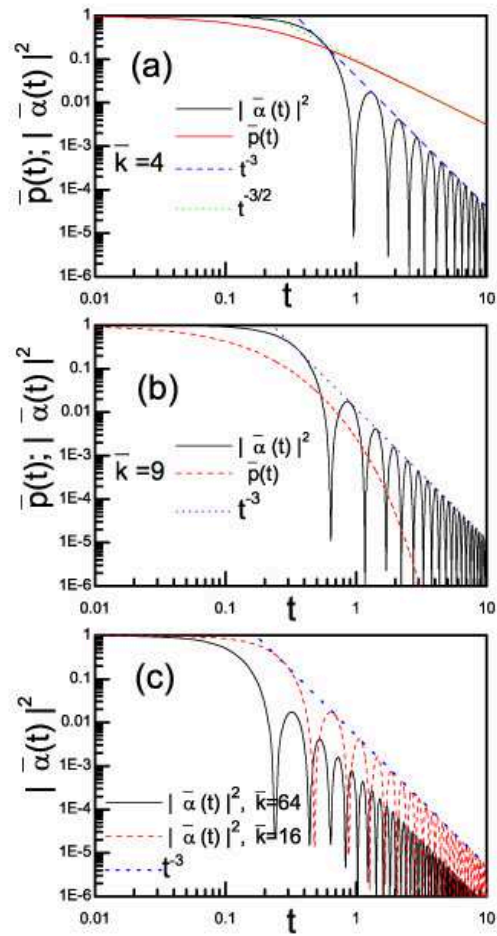}} \caption{(Color
online) Efficiency of the classical and quantum transport. (a)Time
evolution of $\bar{p}(t)$ and $|\bar{\alpha}(t)|^2$ for $\bar{k}=4$.
Both the classical and quantum transport at long time scales show
the power law behavior. The exponent of local maxima of
$|\bar{\alpha}(t)|^2$ is twice the exponent of $\bar{p}(t)$,
implying the quantum transport is more efficient than the classical
one. (b)Time evolution of $\bar{p}(t)$ and $|\bar{\alpha}(t)|^2$ for
$\bar{k}=9$. Here, $\bar{p}(t)$ drops below $|\bar{\alpha}(t)|^2$
and shows an exponential decay. Thus for $\bar{k}=9$ the classical
transport is more efficient than the quantum one.
(c)$|\bar{\alpha}(t)|^2$ versus $t$ for $\bar{k}=64$ and
$\bar{k}=16$. The curve depresses vertically on highly connected
networks, suggesting that the quantum transport becomes more
efficient when the connectivity of network increases.
 \label{fg2}}
\end{figure}
\section{Efficiency of transport on infinite networks}
On finite networks, both the classical and quantum return
probabilities do not decay ad infinitum but reach a constant value
at some time~\cite{rn29}. This value is related to the size of the
networks. To reveal the decay behavior at large time scales, we
consider $<\bar{p}(t)>$ and $<\bar{\pi}(t)>$ on infinite networks.
In this case, the spectrum density can be regarded as a continuous
distribution. Because the networks considered here are uncorrelated
random networks~\cite{rn31}, the spectral density of the Laplacian
Matrix converges to the semicircular distribution,
\begin{equation}\label{eq7}
 \rho (E)=\left\{
\begin{array}{ll}
\frac{\sqrt{4\sigma^2-(E-\bar{k})^2}}{2\pi \sigma^2},   & {\rm if} \ |E -\bar{k}|<2\sigma ,\\
0,    & Otherwise.
\end{array}
\right.
\end{equation}
Where $\sigma=\sqrt{Np(1-p)}$ and $\bar{k}=p(N-1)$ for the ER
networks~\cite{rn32}. This theorem is also known as Wigner¡¯s
law~\cite{rn33}, and it has extensive applications in statistical
physics, solid-state physics and complex quantum-mechanical
systems~\cite{rn34,rn35}.

For sparse networks, i.e., $p<<1$, $\sigma^2$ can be simplified as
$\sigma^2\approx \bar{k}$. Thus Eq.~(\ref{eq7}) can be written as,
\begin{equation}\label{eq8}
 \rho (E)=\left\{
\begin{array}{ll}
\frac{\sqrt{4\bar{k}-(E-\bar{k})^2}}{2\pi \bar{k}},   & {\rm if} \ |E -\bar{k}|<2\sqrt{\bar{k}} ,\\
0,    & Otherwise.
\end{array}
\right.
\end{equation}
Therefore the spectral density is only a function of the average
degree $\bar{k}$ on large sparsely connected networks. Using the
above expression, we can calculate the return probabilities of
Eqs.~(\ref{eq3}) and (\ref{eq4}) in the continuum limit as follows,
\begin{equation}\label{eq9}
\bar{p}(t)=\int e^{-tE}\rho (E)dE=\int
\frac{e^{-tE}\sqrt{4\bar{k}-(E-\bar{k})^2}}{2\pi \bar{k}}dE,
\end{equation}
and
\begin{equation}\label{eq10}
\begin{array}{ll}
\bar{\pi}(t)&\geqslant |\frac{1}{N}\sum_n e^{-itE_n}|^2\equiv
|\bar{\alpha}(t)|^2,\\
|\bar{\alpha}(t)|^2 &= |\int e^{-itE}\rho (E)dE|^2\\
&=|\int \frac{e^{-itE}\sqrt{4\bar{k}-(E-\bar{k})^2}}{2\pi
\bar{k}}dE|^2
\end{array}
\end{equation}
Where the lower bound is obtained by using the Cauchy-Schwarz
inequality and exact for regular networks~\cite{rn26}. Analogous to
the classical case, $|\bar{\alpha}(t)|^2$ depends only on the
eigenvalues of the Hamiltonian. Although $|\bar{\alpha}(t)|^2$ is a
lower bound and differs from the exact value $\bar{\pi}(t)$,
$|\bar{\alpha}(t)|^2$ quantitatively reproduce the overall behavior
of $\bar{\pi}(t)$~\cite{rn29}. Therefore it is appropriate to use
the decay of $|\bar{\alpha}(t)|^2$ to measure the efficiency of the
quantum transport~\cite{rn29}.

Fig.~\ref{fg2} shows the temporal behavior of $\bar{p}(t)$ and
$|\bar{\alpha}(t)|^2$ according to Eqs.~(\ref{eq9}) and (\ref{eq10})
for different values of average degree $\bar{k}$. For $\bar{k}=4$
(Fig.~\ref{fg1}(a)), $\bar{p}(t)$ scales as $\bar{p}(t)\sim
t^{-1.5}$ and the local maxima of $|\bar{\alpha}(t)|^2$ scales as
$|\bar{\alpha}(t)|^2\sim t^{-3}$ at long times. Such scaling
argument can be understood by the spectral density. When
$\bar{k}=4$, the spectral density becomes as $\rho
(E)=\sqrt{8E-E^2}/8\pi$. The long time behavior of $\bar{p}(t)$ and
$|\bar{\alpha}(t)|^2$ are mainly determined by small $E$ values,
thus we can assume $\rho (E)\sim \sqrt{E}$~\cite{rn29}. Such scaling
behavior leads to the power law behavior of the return
probabilities, where the exponent for the quantum transport is twice
the exponent of its classical counterpart~\cite{rn29}.

The behavior of $\bar{p}(t)$ and $|\bar{\alpha}(t)|^2$ alters
accordingly when the average degree $\bar{k}$ increases. In
Fig.~\ref{fg2}(b), we show $\bar{p}(t)$ and $|\bar{\alpha}(t)|^2$
for $\bar{k}=9$. Now the classical $\bar{p}(t)$ does not show
scaling but displays an exponential decay. We note that $\bar{p}(t)$
decays faster than $|\bar{\alpha}(t)|^2$, this indicates that the
classical excitation spreads over the network faster than the
quantum one. As a matter of fact, we find that for networks with
$\bar{k}>4$, $\bar{p}(t)$ always shows a faster decay at long times
compared to $|\bar{\alpha}(t)|^2$. Such a behavior indicates that
the classical transport is more efficient than the quantum transport
on networks with large average degree.
\begin{figure}
\scalebox{1.0}[0.9]{\includegraphics{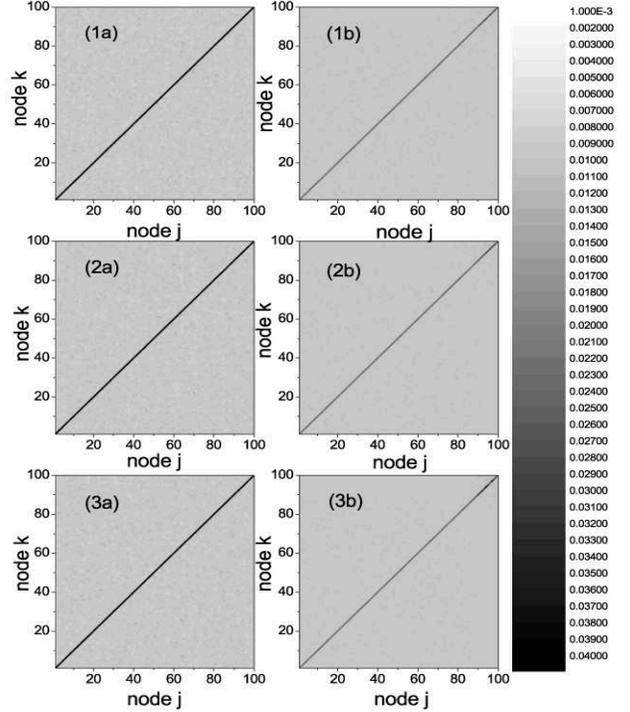}}
 \caption{(Coloronline)
 Ensemble averaged transition probability $<\chi_{k,j}>$ for random
 networks generated by the ER model (column (a)) and configuration
 model (column (b)) with different values of degree $10$, $20$ and
 $30$ (rows 1-3). The network size is $N=100$ and the average is
 over $100$ realizations. The colormap is shown in the right hand of the plot.
 Dark regions denote large values of $<\chi_{k,j}>$ and bright
 regions low values of $<\chi_{k,j}>$.
\label{fg3}}
\end{figure}

The behavior $|\bar{\alpha}(t)|^2$ is qualitatively the same when
the average degree $\bar{k}$ increases. To incarnate the difference
of $|\bar{\alpha}(t)|^2$, we plot $|\bar{\alpha}(t)|^2$ as a
function of $t$ for $\bar{k}=16$ and $\bar{k}=64$ in
Fig.~\ref{fg2}(c). As we can see, the local maxima of
$|\bar{\alpha}(t)|^2$ scales as $|\bar{\alpha}(t)|^2\sim t^{-3}$ for
both the values of $\bar{k}$. However, the scaling depresses
vertically for large $\bar{k}$. This vertical depression of
$|\bar{\alpha}(t)|^2$ suggests that the quantum transport becomes
more efficient when the network connectivity increases.

We note that the increase of average degree greatly changes the
efficiency of the classical transport. This is ascribed to large
value of the eigenvalues $E$ for network with large average degree.
For large $\bar{k}$, $\bar{p}(t)$ decays exponentially, in contrast
to the power law decay for small $\bar{k}$. Quantum mechanically,
the increase of average degree does not change the power law
behavior of the local maxima of $|\bar{\alpha}(t)|^2$, but the first
minimum is reached quickly and the local maxima of
$|\bar{\alpha}(t)|^2$ become lower on highly connected networks.
\begin{figure}
\scalebox{1.0}[1.0]{\includegraphics{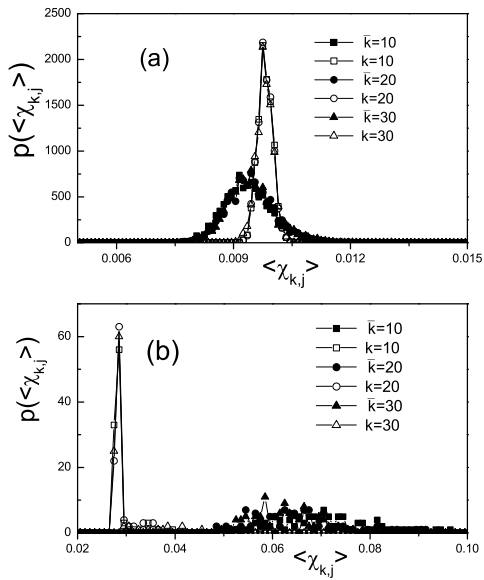}} \caption{
Distribution of ensemble averaged transition probability
$<\chi_{k,j}>$ in the region $0.005<<\chi_{k,j}><0.015$ (a) and
$0.02<<\chi_{k,j}><0.1$ (b). The $k$ with and without bar presents
the random graphs generated by the ER model and configuration model,
respectively. The peaks in (a) are from the non-diagonal transition
probabilities and the peaks in (b) corresponds to the return
probabilities. \label{fg4}}
\end{figure}
\section{Long time averages}
On finite networks, the transition probability converges to a
certain value, this value is determined by the long time average.
Classically, the long time averaged transition probability equals to
the equal-partitioned probability $1/N$. However, the quantum
mechanical transport does not lead to equipartition. Taking into
account the ensemble average, we have,
\begin{equation}\label{eq11}
\begin{array}{ll}
<\chi_{k,j}>&=<(\lim_{T\rightarrow \infty}\frac{1}{T}\int_0^T
\pi_{k,j}(t)dt)>\\
&=<(\sum_{n,l}<k|q_n><q_n|j><k|q_l><q_l|j> \\
&\  \ \  \ \times\lim_{T\rightarrow \infty}\frac{1}{T}\int_0^T e^{-it(E_n-E_l)}dt) >\\
&=<(\sum_{n,l}\delta (E_n-E_l)<k|q_n><q_n|j>\\
&\  \ \  \ \times <k|q_l><q_l|j>) >.
\end{array}
\end{equation}
Where $\delta (E_n-E_l)=1$ for $E_n=E_l$ and $\delta (E_n-E_l)=0$
else. Here, we numerically calculate the ensemble averaged
transition probabilities according to the above equation. The
results are show in Fig.~\ref{fg3}. The dark regions denote large
values of $<\chi_{k,j}>$ and bright regions low values of
$<\chi_{k,j}>$. The dark blocks in the diagonal positions
corresponds to the large return probability. To view the
quantitative behavior of $<\chi_{k,j}>$, we plot the distribution of
$<\chi_{k,j}>$ between all the pairs of two nodes in Fig.~\ref{fg4}.
The peaks in Fig.~\ref{fg4}(a) correspond to the non-diagonal
transition probabilities while the peaks in Fig.~\ref{fg4}(a)
correspond to the return probabilities~\cite{rn13}. The solid
symbols denote the results for the ER networks, and the hollow
symbols denote the numerical results for the random graphs generated
by the configuration model where the degrees are exactly equal to
the average degree of ER networks. It is observed that mean value of
the non-diagonal transition probability for the ER model is smaller
than that for the configuration model (see Fig.~\ref{fg4}(a)). A
contrary conclusion for the return probability is demonstrated in
Fig.~\ref{fg4}(b). We also find that the increase of degree does not
change the central values of the peaks (compare the different
symbols in the figure). This suggests that the quantum transition
probabilities may do not change greatly when connectivity of the
networks increases~\cite{rn13}.
\begin{figure}
\scalebox{0.6}[0.5]{\includegraphics{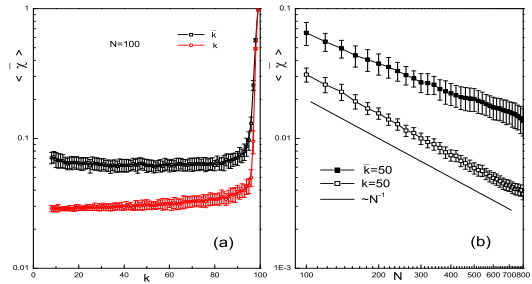}} \caption{ Average
return probability $<\bar{\chi}>$ on random networks generated by
the ER model (marked by $\bar{k}$) and configuration model (marked
by $k$). (a)$<\bar{\chi}>$ versus $k$ on networks of size $N=100$.
(b)$<\bar{\chi}>$ versus network size $N$ when the degree equals to
$50$.
 \label{fg5} }
\end{figure}
In order to reveal how the return probability is related to the
average degree, we show both the node and ensemble averaged return
probability ($<\bar{\chi}>=<\sum_j\chi_{j,j}/N>$) in
Fig.~\ref{fg5}(a). As we can see, the average return probability
$<\bar{\chi}>$ is almost a constant value in a wide range of average
degree $\bar{k}$ but increases drastically when the network
approaches to be fully connected~\cite{rn13}. Here, the difference
for the two random graph models is also visible. The saturated value
$<\bar{\chi}>$ of the ER networks is larger than that of the random
graphs generated by the configuration model. In addition, we have
also studied the relationship of the average return probability
$<\bar{\chi}>$ and the network size $N$, which is shown in
Fig.~\ref{fg5}(b). It is found that the averaged return probability
$<\bar{\chi}>$ is inversely proportional to the network size $N$
(see the asymptote indicated in the figure). This means that the
return probability of small-size networks is larger than that of
large-size networks.
\section{Transport on extremely connected networks}
As we have shown, the average return probability $<\bar{\chi}>$
increases rapidly when the networks approach to be fully connected.
To study this issue in detail, we consider the transport on
extremely connected networks. In such case, the connection
probability $p$ approaches to $1$ and the degree of the network is
of the order ${\cal O}(N)$. We construct the extremely connected
networks by randomly removing a certain number of edges on the fully
connected network. The resulting network is equivalent to the ER
random network, and the algebra of removing edges is much
computationally cheaper than the ER random graph algebra.

\begin{figure}
\scalebox{1.0}[1.0]{\includegraphics{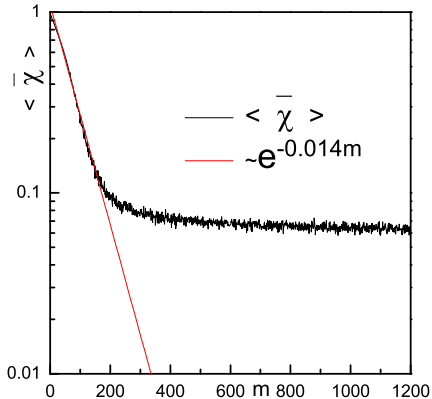}}
 \caption{(Color
online) Average return probability $<\bar{\chi}>$ versus the number
of removed edges $m$ on networks of size $N=100$. The straight line
indicates the exponential decay for $m<200$.
\label{fg6}}
\end{figure}

Here, we consider the transport on networks of size $N=100$, and
assume the number of removed edges is $m$. Fig.~\ref{fg6} shows the
average return probability $<\bar{\chi}>$ as a function of the
number of removed edges. We note that $<\bar{\chi}>$ decreases to
$\sim 0.1$ when only 4\% edges ($200$ edges) are removed. As more
and more edges removed, $<\bar{\chi}>$ decreases slowly and tends to
reach the saturated value $\sim 0.65$. We find that the
$<\bar{\chi}>$ versus $m$ can be well described by an exponential
decay $<\bar{\chi}>\sim e^{-0.014m}$ in the region $m<200$. When
only a few edges are removed, $<\bar{\chi}>$ is close to the return
probability of the complete network (or fully connected network).
For a fully connected network of size $N$, one eigenvalue of the
Hamiltonian is $0$ and all the other values are equal to $N$, the
long time averaged return probability is
$\bar{\chi}=(N^2-2N+2)/N^2$. This is a striking feature of CTQWs
which differs from the classical counterpart.
\section{Conclusions and Discussions}
In conclusion, we have studied the classical and quantum transport
on Erd\"{o}s-R\'{e}nyi networks. We numerically calculate the
ensemble averaged transition probability between two nodes of the
networks. For finite networks, we find that the limiting transition
probability is reached very quickly. For infinite networks whose
spectral density follows the semicircle law, the efficiencies of the
classical and quantum mechanical transport are compared on networks
of different average degree $\bar{k}$. It is shown that the
classical transport is more efficient than the quantum transport on
networks with large connectivity. In the long time limiting, we
consider the distribution of the ensemble averaged transition
probability, and show that the quantum transport exhibits a high
return probability. Such high return probability almost do not
change in a wide range of connection probability $p$ but increases
rapidly when the network approaches to be fully connected. For
networks whose topology is not extremely connected, the return
probability is inversely proportional to the network size $N$. In
addition, we also compare the results with that on a random graph
model in which the degree of each node equals to the average degree
$\bar{k}$ of the Erd\"{o}s-R\'{e}nyi networks.

We have shown that the transport dynamics on the ER networks is
different from that on the equivalent network generated by the
configuration model. The difference may be related to the different
ensembles of the network model. Since the return probability
reflects the symmetry of the network structure, the high return
probability on ER networks may suggest a high symmetry of the
topology~\cite{rn36}.
\begin{acknowledgments}
The authors would like to thank Zhu Kai for converting the
mathematical package used in the calculations. This work is
supported by the Cai Xu Foundation for Research and Creation (CFRC),
National Natural Science Foundation of China under project 10575042
and MOE of China under contract number IRT0624 (CCNU).
\end{acknowledgments}


\begin{thebibliography} {Albert2000}
\bibitem{rn1} J. Feldmann, et al., Phys. Rev. Lett. {\bf 70}, 3027 (1993).
\bibitem{rn2} T. Stroucken, A. Knorr, P. Thomas, and S. W. Koch, Phys. Rev. B {\bf 53}, 2026(1996).
\bibitem{rn3} M. Dyakonov, et al., Phys. Rev. B {\bf 56}, 10412 (1997)
\bibitem{rn4} G. R. Allan and H. M. van Driel, Phys. Rev. B {\bf 59}, 15740 (1999).
\bibitem{rn5} N. Ashwin and V. Ashvin, quant-ph/0010117.
\bibitem{rn6} G. Abal, R. Siri, A. Romanelli, et al., Phys. Rev. A {\bf 73}, 042302(2006).
\bibitem{rn7} D. Solenov and L. Fedichkin, Phys. Rev. A {\bf 73}, 012313(2003).
\bibitem{rn8} F. Sorrentino, M. di Bernardo, G. H. Cu\'ellar, and S. Boccaletti, Physica D {\bf 224}, 123 (2006).
\bibitem{rn9} H. Krovi and T. A. Brun, Phys. Rev. A {\bf 73}, 032341 (2006).
\bibitem{rn10} O. M\"{u}lken and A. Blumen, Phys. Rev. E {\bf 71}, 016101 (2005).
\bibitem{rn11} O. M\"{u}lken, V. Bierbaum and A. Blumen, J. Chem. Phys {\bf 124}, 124905 (2006).
\bibitem{rn12} W. Barford and C. D. P. Duffy, Phys. Rev. B {\bf 74} 075207 (2006).
\bibitem{rn13} O. M\"{u}lken, V. Pernice and A. Blumen, Phys. Rev. E {\bf 76}, 051125 (2007).
\bibitem{rn14} Y. Aharonov, L. Davidovich, and N. Zagury, Phys. Rev. A {\bf 48}, 1687 (1993).
\bibitem{rn15} N. Shenvi, J. Kempe, and K. Brigitta Whaley, Phys. Rev. A {\bf 67}, 052307 (2003).
\bibitem{rn16} P. Erd\"{o}s and A. R\'{e}nyi, Publicationes Mathematicae {\bf 6}, 290 (1959)
\bibitem{rn17} P. Erd\"{o}s and A. R\'{e}nyi, Publications of the Mathematical Institute of the Hungarian Academy of Sciences {\bf 5}, 17(1960).
\bibitem{rn18} P. Erd\"{o}s and A. R\'{e}nyi, Acta Mathematica Scientia Hungary {\bf 12}, 26 (1961).
\bibitem{rn19} M. E. J. Newman, cond-mat/0202208.
\bibitem{rn20} R. Albert and A.-L. Barab\"{a}si, Rev. Mod. Phys 74, 47 (2002).
\bibitem{rn21} M. E. J. Newman, SIAM Review, {\bf 45}, 167 (2003).
\bibitem{rn22} R. Milo et al., Science, {\bf 303}, 1538 (2004).
\bibitem{rn23} B. J. Kim et al., Phys. ReV. E {\bf 69}, 045101 (2004).
\bibitem{rn24} S. Maslov and K. Sneppen, Science {\bf 296}, 910 (2002).
\bibitem{rn25} M. E. J. Newman, Phys. Rev. Lett {\bf 95}, 108701 (2005).
\bibitem{rn26} A. Volta, O. M\"{u}lken and A. Blumen, J. Phys. A {\bf 39}, 14997 (2006).
\bibitem{rn27} O. M\"{u}lken, A. Volta and A. Blumen, Phys. Rev. A {\bf 72}, 042334 (2005).
\bibitem{rn28} E. Farhi and S. Gutmann, Phys. Rev. A {\bf 58}, 915 (1998).
\bibitem{rn29} O. M\"{u}lken and A. Blumen, Phys. Rev. E {\bf 73}, 066117 (2006).
\bibitem{rn30} O. M\"{u}lken and A. Blumen, Phys. Rev. A {\bf 73}, 012105 (2006).
\bibitem{rn31} I. J. Farkas et al., Phys. Rev. E {\bf 64}, 026704 (2001).
\bibitem{rn32} B. Gong, L. Yang and K. Yang, Phys. Rev. E {\bf 72}, 037101 (2005).
\bibitem{rn33} E. P. Wigner, Ann. Math {\bf 62}, 548 (1955); {\bf 65}, 203 (1957); {\bf 67}, 325 (1958).
\bibitem{rn34} M. L. Mehta, \emph{Random Matrices} (2nd ed, Academic, New York, 1991).
\bibitem{rn35} A. Crisanti, G. Paladin and A. Vulpiani, \emph{Products of Random Matrices in Statistical Physics} (Springer Series in Solid-State
Sciences Vol. 104 Springer, Berlin, 1993).
\bibitem{rn36} Y. Xiao et al., arXiv:0709.1249.
\end{thebibliography}
\end{document}